%% file: main.tex
\documentclass[dvipsnames,format=sigconf,anonymous=false,review=false]{acmart}

\AtBeginDocument{%
  }

\copyrightyear{2026}
\acmYear{2026}
\setcopyright{cc}
\setcctype{by}
\acmConference[GECCO Companion '26]{Genetic and Evolutionary Computation Conference}{July 13--17, 2026}{San Jose, Costa Rica}
\acmBooktitle{Genetic and Evolutionary Computation Conference (GECCO Companion '26), July 13--17, 2026, San Jose, Costa Rica}
\acmDOI{10.1145/3795101.3814699}
\acmISBN{979-8-4007-2488-6/2026/07}

\usepackage{algorithm}
\usepackage{algpseudocode}
\usepackage{graphicx}
\usepackage{tikz}
\usepackage{subcaption}
\usepackage{amsmath}
\usepackage{booktabs}
\usepackage{multirow}
\usepackage{fontenc}
\usepackage{makecell}
\tikzset{
    city/.style={circle, draw=black, fill=gray!20, minimum size=6mm, inner sep=0pt, font=\small\sffamily},
    supernode/.style={ellipse, draw=black!70, fill=gray!5, minimum height=1.2cm, minimum width=2.2cm, font=\small\sffamily, align=center},
    logical/.style={circle, draw=black, minimum size=1cm, inner sep=0pt, font=\normalsize\bfseries\sffamily},
    fixed/.style={ultra thick, blue!70!black},
    otherfixed/.style={ultra thick, red!70!black},
    potential/.style={thick, gray!50},
    endpoint/.style={font=\scriptsize\itshape}
}

\begin{document}

\title{A Hybrid Classical-Quantum Annealing Algorithm for the TSP}
\author{Siwei Hu}
\orcid{0009-0009-1040-8839}
\affiliation{%
  \institution{Dipartimento di Informatica, Università di Roma “La Sapienza”}
  \city{Rome}
  \country{Italy}}
\email{hu.1985972@studenti.uniroma1.it}

\author{Victor Lopata}
\orcid{0009-0007-3353-1325}
\affiliation{%
  \institution{Dipartimento di Informatica, Università di Roma “La Sapienza”}
  \city{Rome}
  \country{Italy}}
\email{lopata.2005708@studenti.uniroma1.it}

\author{Salvatore Sinno}
\orcid{0009-0002-9177-5161}
\affiliation{%
  \institution{Advanced Research \& Innovation Group, Unisys}
  \city{Milton Keynes}
  \country{UK}
  }
\email{salvatore.sinno@unisys.com}

\author{Shruthi Thuravakkath}
\orcid{0009-0002-9792-3276}
\affiliation{%
  \institution{Advanced Research \& Innovation Group, Unisys}
  \city{Bangalore}
  \country{India}}
\email{Shruthi.Thuravakkath@unisys.com}

\author{Paolo Zuliani}
\orcid{0000-0001-6033-5919}
\affiliation{%
  \institution{Dipartimento di Informatica, Università di Roma “La Sapienza”}
  \city{Rome}
  \country{Italy}}
\email{zuliani@di.uniroma1.it}


\input{Sections/Abstract}


\maketitle


\input{Sections/Introduction}
\input{Sections/Related_Work}

\input{Sections/Background}
\input{Sections/ProposedHybridMethodology}

\input{Sections/ExperimentalSetup}

\input{Sections/Results}

\input{Sections/Conclusions}

\begin{acks}
    S.H. and V.L. were supported by a Unisys UK grant to Sapienza University. S.H. and P.Z. were partly supported by the Sapienza University project D2QNeT (RG1241910FF320FB). Access to the D-Wave platform was provided by CINECA (Italy).
\end{acks}


\bibliographystyle{ACM-Reference-Format}
\bibliography{Sections/Biblio}


\end{document}

%% file: Sections/Abstract.tex
\begin{abstract}
Hybrid quantum-classical algorithms can help mitigating the physical limitations of current quantum devices, particularly the low qubit count and the reduced topological connectivity. In this paper, we propose a hybrid technique to solve a well-known NP-hard optimization problem: the Traveling Salesperson Problem (TSP). Our approach is based on a graph contraction technique that removes most of the dimensionality of the original problem instance, producing a sub-TSP of a size suitable to be efficiently solved by a quantum device. The performance of our approach is first demonstrated on classical quantum simulation using Path Integral Monte Carlo, and then run on a D-Wave quantum annealer.
\end{abstract}

%% file: Sections/Introduction.tex
\section{Introduction}\label{sec:intro}

The Traveling Salesperson Problem \cite{Gavish_Graves_1978, TSP_1956}, known as TSP, is probably one of the most studied NP-hard problems in theoretical computer science and combinatorial optimization. Given a set of $N$ cities, the problem consists in determining the shortest possible route to visit each city, starting from and returning to a specific one. In terms of graph theory, given a complete graph of $N$ nodes and a weight function that represents the distances between the cities, the objective of the problem is to find the shortest Hamiltonian cycle. The scientific literature offers several algorithmic approaches to solving the problem, such as exact algorithms ({\em e.g.} Branch and Cut methods) for which solvers, like Concorde, have successfully solved instances exceeding 85000 nodes \cite{APPLEGATE200911}. Other approaches rely on heuristics (such as the Lin-Kernighan heuristic \cite{HELSGAUN2000106}) or nature-inspired metaheuristics \cite{585892} that focus on reducing computational complexity, albeit producing suboptimal solutions.

Over the last decade, there has been a particular interest in the use of quantum approaches to solve optimization problems. 
Thanks to the emergence of quantum annealers like D-Wave's, and neutral atom systems like QuERA and PASQAL, gate-based systems from tech giants like IBM \cite{IBM_2025}, Google, and specialized companies like IonQ and Quantinuum, quantum computing hardware has become more accessible. However, despite these developments, the existing approaches on all these systems are still limited in terms of scalability in the number of qubits and their connectivity. Classical simulations and hybrid classical-quantum algorithms can be a useful approach to overcome the physical limitations of current Noisy Intermediate-Scale Quantum (NISQ) \cite{NISQ} devices. 

Our technique to solve the TSP is based on the classical simulation of Quantum Annealing (QA) \cite{KadowakiQA} via the Path Integral Monte Carlo (PIMC) method \cite{QA_2002}. In particular, Martoňák et al. \cite{QA_2004} proposed a PIMC quantum annealing scheme based on a highly constrained Ising-like representation of the TSP. While their theoretical framework demonstrates the promise of quantum annealing for the TSP, its direct translation to physical hardware (as well as its large-scale classical simulation) is a complicated task, primarily due to the massive qubit requirements and the lack of sufficient inter-qubit connectivity in current quantum devices. Our proposed hybrid approach effectively resolves this bottleneck. By leveraging classical optimization to contract the graph, we not only achieve a highly scalable PIMC simulation, but we also reduce the instance to a sufficiently small sub-TSP, making it a candidate to be successfully embedded in NISQ annealers. 
As will be shown, our technique can also be adapted to quantum hardware. In particular, we will analyze how this approach interacts with a D-Wave quantum annealer. 

The paper is organized as follows: Section \ref{sec:relwork} presents related works for solving the TSP
; Section \ref{sec:Background} explains the fundamental theory behind our approach; Section \ref{sec:prop} introduces the proposed hybrid approach; Section \ref{sec:expsetup} and \ref{sec:results} illustrate the experimental setup and the results. Finally, Section 7 concludes the paper. 

%% file: Sections/Related_Work.tex
\section{Related Work}\label{sec:relwork}

Since the TSP is NP-hard, no known deterministic algorithm can solve all instances in polynomial time. 
Classically, the TSP has been extensively studied through both exact methods and heuristic/metaheuristic approaches \cite{Darrel1, Darrell2, Darrell3}. In practice, however, advanced heuristics often dominate for large-scale instances. General purpose metaheuristics such as Simulated Annealing (SA) \cite{SA_kirkpatrick} have provided a conceptual framework that has also been combined with local search to improve both exploration and intensification \cite{SA_johnson, SA_martin}. 

In recent decades, quantum and quantum-inspired approaches have also been proposed \cite{QuantumOpt}, such as Quantum Annealing and universal gate-based approaches like the Quantum Approximate Optimization Algorithm (QAOA) \cite{QAQA_farhi, QAQA_hadfield} and Quantum Phase Estimation (QPE). While QAOA and QPE offer theoretical advantages, their scaling requirements coupled with deep circuit requirements and sensitivity to problem density, render them exceedingly difficult to implement for large scale TSP instances in the NISQ era \cite{TSP_differentArch, VariationalAlg_2025, NISQ}. Consequently, QA has emerged as a potentially better alternative for near-term applications. The paradigms of Adiabatic Quantum Computation \cite{AQC_Albash_2018, AQC_2014book} and Quantum Annealing \cite{QA_2011Jonson, QA_kim} introduce quantum fluctuations as an alternative mechanism to thermal noise in SA. 
Before physical quantum annealers achieved sufficient scale, Marto\v{n}\'ak et al. \cite{QA_2004, QA_2002} proposed a seminal Path-Integral Monte Carlo (PIMC) scheme to simulate quantum annealing for the TSP, by mapping it to a highly constrained Ising-like spin system, where valid tours correspond to specific spin configurations. The approach applies a PIMC quantum annealing scheme using 2-opt moves as basic updates. Evaluated against the \texttt{pr1002} TSPLIB \cite{tsplib} instance (1002 cities), this PIMC QA scheme \cite{QA_2004} demonstrated superior accuracy compared to classical thermal SA, proving that quantum tunneling mechanisms could efficiently navigate local minima in the complex TSP landscape. 

With respect to QA physical implementation, D-Wave provides their quantum annealer and development kit Ocean that allow solving computational optimization problems using the QUBO model \cite{DWave_jobshp}. Such machines have already been exploited to solve the TSP in recent years \cite{TSP_ApplicationDriven, TSP_Dwave2021, TSP_EnhancedQA2025, TSP_differentArch}. However one major concern that emerged is the minor embedding problem \cite{MinorEm_choi, MinorEm_spagnoli}: fully-connected TSP graphs quickly exhaust the sparse connectivity limits of current Quantum Processing Units (QPUs). It has been shown that the current embedding schemes struggle when the number of cities exceeds 15 \cite{TSP_ApplicationDriven}. To bridge this gap, decomposition methods have been employed in order to partition large QUBO problems into smaller sub-problems. D-Wave's former partitioning tool \texttt{qbsolv} \cite{Booth2017PartitioningOP}, relied on \textit{impact} values defined as how much the current solution varies once the variables chosen are changed. However, empirical results show that it often extracts highly correlated, fixed structures that trap the hybrid solver in local optima. To rectify this behavior, Atobe et al. \cite{Subqubo} proposed a rigorous hybrid annealing method based on an iterative subQUBO model extraction. In our work, we instead synthesize two foundational methodologies: by implementing the PIMC approach modeled in \cite{QA_2004}, we isolate the quantum tunneling dynamics required to escape classical TSP valleys, then we extend this framework into the subQUBO domain outlined in \cite{Subqubo}.

%% file: Sections/Background.tex
\section{Background}\label{sec:Background}
In this section we briefly present the main components that support our work. 

\subsection{TSP and Quantum Annealing}
The TSP can be stated on a complete weighted graph $G = (V,E)$, where $V$ is a set of cities such that $|V|= N $, and a distance matrix $D=[d_{i,j}], \space (i,j) \in E$ specifies the cost of traveling between each pair of nodes $i$ and $j$. The objective is to identify an optimal Hamiltonian cycle that visits every node exactly once and returns to the origin, such that the total distance or cost is minimized. Mathematically, this involves finding a permutation $\pi$ of the cities that minimizes the function:

\[ \left(\sum_{i=1}^{N-1}{d_{\pi(i),\pi(i+1)}}\right)+d_{\pi(N),\pi(1)}\space.\]

In the adiabatic quantum computation paradigm \cite{AQC_2014book}, we consider a time-dependent Hamiltonian $H(t)$ that interpolates from an initial Hamiltonian $H_0$ to a problem Hamiltonian $H_p$. Conventionally $H_0$ is chosen with an easy to prepare minimal energy configuration, then by gradually changing the system parameters, the system transitions from $H_0$ to the lowest energy state of $H_p$, which encodes the solution of the problem. If such transition is done sufficiently slow, the system is likely to remain in the ground state, thus allowing to identify the optimal solution of a given combinatorial optimization problem. Formally, the annealing process can be described by:
\[H(t) = s(t)H_0 + (1-s(t))H_p\]
where $s(t)\in [0,1]$, and the annealing process starts with $s(t) = 1$ and evolves towards $s(t) =0$.

In the context of physical devices such as those produced by D-Wave, the Hamiltonian is constrained to take the form of an Ising model, a statistical mechanics representation of interacting spins \cite{Ising_2014, AQC_2014book}. Mathematically the Ising model is shown to be equivalent to Quadratic Unconstrained Binary Optimization (QUBO) models \cite{QUBO_tutorial}, which seek to minimize $y = x^tQx$ for binary variables $x \in \{0,1\}^n$. That equivalence allows combinatorial optimization problems to be solved via quantum annealing by mapping their objective functions and constraints into the coefficients of the $Q$ matrix, which serves as the input for the QA hardware. To cast the TSP into the QUBO framework we follow  \cite{TSP_EnhancedQA2025}, which leverages a binary indicator matrix $X$ of size $N \times N$, where the variable $x_{i,j}$ takes the value 1 if city $i$ is visited at time step $j$, and 0 otherwise. Then the goal is to minimize the tour distance:
\[
C_{\textit{dist}} = \sum_{j=0}^{N-1}\sum_{i=0}^{N-1}\sum_{k=0}^{N-1}d_{i,k}x_{i,j}x_{k,((j+1) \space \text{ mod } \space N)} \space. 
\] In order to find a valid tour, several constraints have to be taken into account. $P_{\textit{city}}$ enforces every city to be visited exactly once; $P_\textit{time}$ states that at each discrete time step, exactly one city must be visited. These two constraints can be respectively encoded as following:
\[
P_{\textit{city}} =\lambda_1 \sum_{i=0}^{N-1}\left (\sum_{j=0}^{N-1}x_{i,j}-1\right)^2 \quad \quad P_{\textit{time}} =\lambda_2 \sum_{j=0}^{N-1}\left (\sum_{i=0}^{N-1}x_{i,j}-1\right)^2 \space .
\]

The choice of the Lagrange multipliers $\lambda_1$ and $\lambda_2$ is critical: they must be set high enough to preclude constraint violations, but not too high that they compress the energy differences between valid tours, which would render the optimal solution indistinguishable from near-optimal configurations. As a refinement, Alawir et al. \cite{TSP_EnhancedQA2025} proposed to fix  the start tour at city $0$, meaning that $x_{0,0}$ is always 1, which would reduce the number of variables to $(N-1)^2$: since any Hamiltonian cycle can start at an arbitrary node without loss of optimality, fixing city $0$ at time $0$ eliminates one row and one column from $X$. The complete refined QUBO formulation is:
\begin{equation}\label{qubo}
    \begin{aligned}
      Q' =  \sum_{j=1}^{N-2}\sum_{i=1}^{N-1}\sum_{k=1}^{N-1}d_{i,k}x_{i,j}x_{k,j+1} + \sum_{i=1}^{N-1}\left(d_{0,i}x_{i,1}+d_{i,0}x_{i,N-1}\right) \\ 
    + \lambda_1 \sum_{i=1}^{N-1}\left (\sum_{j=1}^{N-1}x_{i,j}-1\right)^2 + \lambda_2 \sum_{j=1}^{N-1}\left (\sum_{i=1}^{N-1}x_{i,j}-1\right)^2 \space
    \end{aligned}
\end{equation}
where the second term considers the distance from the first city to the second and the distance from the last city to the first one. Such a position-based QUBO formulation already enforces a complete Hamiltonian cycle, therefore additional subtour elimination constraints are not strictly necessary.

\subsection{Path-Integral Scheme}
Marto\v{n}\'ak et al. \cite{QA_2004} applied quantum fluctuations to the TSP using a path-integral Monte Carlo (PIMC) approach. In PIMC, quantum fluctuations are not simulated directly; instead, the \textit{d}-dimensional transverse-field quantum model is mapped onto a \textit{d+1}-dimensional classical model via Trotter discretization \cite{Trotter}. The system is represented by $P$ classical \textit{replicas} as time slices of the original problem ({\em e.g.} $P$ copies of the TSP tour). The additional dimension represents this {\em imaginary} time, and the replicas are coupled along this dimension. This work sought to demonstrate that quantum tunneling could be more effective than thermal activation for navigating the highly constrained Hilbert space of the symmetric TSP. The crucial point that has been highlighted is how to implement a quantum Hamiltonian $H_{\textit{TSP}} = H_{\textit{pot}} + H_{\textit{kin}}$, where $H_{\textit{pot}}$ represents the classical potential energy of a tour, while $H_{\textit{kin}}$ is the kinetic energy operator that induces quantum fluctuations that transition the system directly between valid tours, effectively restricting the search to the feasible subspace of the problem.
To represent the symmetric TSP, where $d_{i,j} = d_{j,i}$, a symmetric matrix $\hat{U} = \hat{T}+ \hat{T}^t$ has been introduced, where $\hat{T}$ is a standard directed permutation matrix. In this formulation $\hat{U}_{i,j} = 1$ if city $i$ and city $j$ are connected, and 0 otherwise. Then the tour length can be expressed as:
\[
H_{\textit{pot}}(\hat{U}) = \frac{1}{2} \sum_{i,j}d_{i,j}\hat{U}_{i,j} \space .
\]
This shift to an undirected representation was motivated by the mechanism of the 2-opt move, chosen to devise a suitable kinetic energy in the problem. The 2-opt-move is a standard heuristic that eliminates two links and rebuilds the tour by reversing an intermediate segment. In a directed representation, reversing a tour segment requires a global reconfiguration of the bit-string, making it difficult to implement in a quantum Hamiltonian. However, in the symmetric $\hat{U}$ matrix, the internal connections remain unchanged, which allows the 2-opt move to be represented as a local four spin-flip operator. 

In order to retain the PIMC simplicity, Marto\v{n}\'ak et al. \cite{QA_2004} made a drastic simplification by replacing the complex four-spin operator with a simpler transverse-Ising-like single spin-flip kinetic term, which is straightforward to Trotter discretize. Crucially, the Monte Carlo moves remain 2-opt move and are restricted to the valid tours subspace, therefore feasibility is preserved by the move set, while quantum effects enter via imaginary time replica coupling. 

\subsection{subQUBO Extraction}

While \cite{QA_2004} focuses on global quantum simulations, the subsequent development of physical Ising machines revealed that the number of available qubits and the density of their connections are often insufficient to host a full TSP instance for non-trivial number of cities. In order to tackle this hardware bottleneck, Atobe et al. \cite{Subqubo} proposed a subQUBO model extraction method that provides a hybrid annealing iterative process that does not need to solve the entire problem on a quantum processor. The core idea is to first identify a subset $N_{var}$ of variables of the current best solution $X$. Then, by extracting a subQUBO $S$ that satisfies $N_{var} \subseteq S$, finding the ground state of $S$ while keeping variables in $X \setminus S$ fixed will yield the global ground-state energy of the original problem.
Since the set $N_{var}$ is unknown, Atobe et al. \cite{Subqubo} introduced a selection heuristic based on the statistical behavior of a pool of solution instances. The algorithm maintains $N_I$ quasi-optimal solutions generated by a classical solver, and  then to extract a subQUBO of size $m$, ideally the hardware limit, $N_S$ instances are selected from the pool and the variability of each binary variable $x_i$ is calculated. More precisely, a variable is considered most varied if its value fluctuates significantly across the sample solutions, which suggests uncertainty regarding its optimal state. Conversely, variables that are stable across all instances are likely already at their optimal value. This method ensures that the limited resources of the quantum annealer are focused on the most unsettled portions of the problem. By repeating this extraction and updating the pool with lower-energy solutions from the QPU, the algorithm navigates toward the ground state of large-scale QUBOs that would not be able to fit on a current quantum annealer.

%% file: Sections/ProposedHybridMethodology.tex
\section{Proposed Hybrid Algorithm}\label{sec:prop}

\begin{figure*}[t]
\centering

\tikzset{
    n/.style={circle, fill=black, inner sep=2.5pt}, 
    fixed/.style={ultra thick, blue!80!black},      
    uncertain/.style={thick, gray!60, dashed},      
    logical/.style={ultra thick, gray!80}           
}

\newcommand{\fullinstance}{
    \node[n] (n1) at (0.0, 3.5) {};
    \node[n] (n2) at (1.0, 4.2) {};
    \node[n] (n3) at (2.2, 4.0) {};
    \node[n] (n5) at (0.5, 2.9) {};
    
    \node[n] (n6) at (4.0, 4.5) {};
    \node[n] (n7) at (5.5, 4.8) {};
    \node[n] (n8) at (6.5, 4.2) {};
    
    \node[n] (n9) at (5.0, 1.0) {};
    \node[n] (n10) at (6.2, 1.5) {};
    
    \node[n] (n11) at (3.0, 2.5) {};
    \node[n] (n12) at (3.5, 0.8) {};
}

\newcommand{\reducedinstance}{
    \node[n] (n3) at (2.2, 4.0) {};  
    \node[n] (n5) at (0.5, 2.9) {};  
    
    \node[n] (n6) at (4.0, 4.5) {};  
    \node[n] (n8) at (6.5, 4.2) {};  
    
    \node[n] (n9) at (5.0, 1.0) {};  
    \node[n] (n10) at (6.2, 1.5) {}; 
    
    \node[n] (n11) at (3.0, 2.5) {}; 
    \node[n] (n12) at (3.5, 0.8) {}; 
}

\newcommand{\chainA}{\draw[fixed] (n3)--(n2)--(n1)--(n5);}
\newcommand{\chainB}{\draw[fixed] (n6)--(n7)--(n8);}
\newcommand{\chainC}{\draw[fixed] (n9)--(n10);}

\newcommand{\logicalbackbone}{
    \draw[logical] (n3)--(n5);
    \draw[logical] (n6)--(n8);
    \draw[logical] (n9)--(n10);
}


\begin{subfigure}[b]{0.32\textwidth}
    \centering
    \begin{tikzpicture}[scale=0.6, every node/.style={scale=0.6}]
        \fullinstance \chainA \chainB \chainC
        \draw[uncertain] (n5)--(n11) (n11)--(n6) (n8)--(n10) (n9)--(n12) (n12)--(n3);
    \end{tikzpicture}
    \caption{Solution 1}
\end{subfigure}
\hfill
\begin{subfigure}[b]{0.32\textwidth}
    \centering
    \begin{tikzpicture}[scale=0.6, every node/.style={scale=0.6}]
        \fullinstance \chainA \chainB \chainC
        \draw[uncertain] (n5)--(n12) (n12)--(n11) (n11)--(n8) (n6)--(n10) (n9)--(n3);
    \end{tikzpicture}
    \caption{Solution 2}
\end{subfigure}
\hfill
\begin{subfigure}[b]{0.32\textwidth}
    \centering
    \begin{tikzpicture}[scale=0.6, every node/.style={scale=0.6}]
        \fullinstance \chainA \chainB
        \draw[uncertain] (n6)--(n12) (n12)--(n10) (n10)--(n11) (n11)--(n9) (n9)--(n5) (n3)--(n8);
    \end{tikzpicture}
    \caption{Solution 3}
\end{subfigure}

\vspace{6mm} 


\begin{center}
\begin{subfigure}[b]{0.45\textwidth}
    \centering
    \begin{tikzpicture}[scale=0.6, every node/.style={scale=0.6}]
        \fullinstance \chainA \chainB \chainC
        
        \node[text=blue!80!black, font=\bfseries\huge] at (1.5, 3.4) {$f=3/3$};
        \node[text=blue!80!black, font=\bfseries\huge] at (5.2, 4.0) {$f=3/3$};
        \node[text=red!80!black, font=\bfseries\huge] at (5.6, 0.5) {$f=2/3 \ge \tau$};
    \end{tikzpicture}
    \caption{Confident edges ($E_{\textit{fixed}}$) evaluated with $\tau = 0.6$}
\end{subfigure}
\hspace{0.05\textwidth} 
\begin{subfigure}[b]{0.45\textwidth}
    \centering
    \begin{tikzpicture}[scale=0.6, every node/.style={scale=0.6}]
        \reducedinstance \logicalbackbone
    \end{tikzpicture}
    \caption{Reduced instance ($N_{\textit{sub}} = 5$)}
\end{subfigure}
\end{center}

\caption{Simplified workflow of the graph contraction process, highlighting the threshold concept. Graphs (a)-(c) depict three TSP solutions sampled from our pool $\mathcal{P}$. The edges highlighted in blue are those that will be fixed. Graph (d) illustrates the fixed edges alongside their respective frequencies. Finally, Graph (e) represents the resulting reduced instance.}
\label{fig: example1}
\end{figure*}
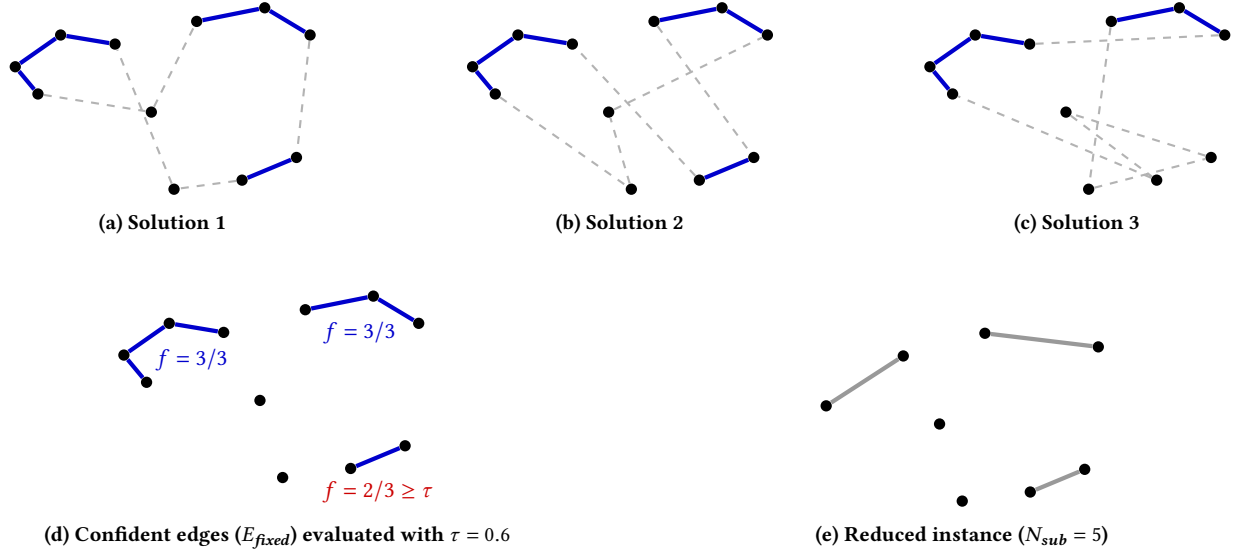

Building upon the subQUBO strategies discussed in the previous section, we propose an approach that is no longer based on the QUBO formulation, but rather on the structure of the graph and its edges, as proposed in \cite{jager2014backbone}. 

Given a TSP stated on a complete graph $G = (V, E)$ and the distance matrix $D$, we initially generate a pool of solutions at a low computational cost with the aim of extracting structural information by identifying the edges that appear most frequently, as it is likely that they belong to the optimal tour. More formally, let \[\mathcal{P} = \{X_1, \hdots, X_{N_I}\}\] be the set of initial TSP solutions. For each edge $e \in E$, we can compute its frequency $\textit{freq}(e)$ within the pool, and construct the set of fixed edges $E_{\textit{fixed}}$ defined as:
\[\textit{freq}(e) = \frac{1}{N_I} \sum_{X \in \mathcal{P}} \mathbb{I}(e \in X), \quad E_{\textit{fixed}} = \{e \in E | \textit{ freq}(e) \geq \tau\}\]
where $\mathbb{I}$ is the indicator function and $\tau \in [0,1]$ a given confidence threshold. 
These edges constitute the ``confident backbone'' of our route, which enables us to contract the graph. Once $E_{\textit{fixed}}$ is computed, we aim to reduce the dimensionality of the original instance by exploiting this set. By design, the fixed edges form a set of disjoint paths (cycles can be easily broken). Once identified, we can easily merge each chain into a single logical entity, which we refer to as a \textit{super-node}. Let $V_{\textit{free}}$ be the set of free cities and let $\mathcal{C}$ be the set of super-nodes. The dimensionality of the reduced problem is given by: 
\[N_{\textit{sub}} = |V_{\textit{free}}| + |\mathcal{C}|\space.\]

To formally define the sub-TSP, we have to compute the new distance matrix $D_{\textit{sub}}$ of size $N_{\textit{sub}} \times N_{\textit{sub}}$. For any super-node $c \in \mathcal{C}$, let $h(c)$ and $t(c)$ be the head and tail of the path, respectively. The elements of the contracted distance matrix $D_{\textit{sub}}$ are calculated in the following way:
\begin{equation}
D_{\textit{sub}}(i, j) = 
\begin{cases} 
d_{ij} & \text{if } i, j \in V_{\textit{free}} \\
\min \big( d_{i, h(j)}, d_{i, t(j)} \big) & \text{if } i \in V_{\textit{free}}, j \in \mathcal{C} \\
\begin{aligned}
    \min \big( & d_{h(i), h(j)}, d_{h(i), t(j)}, \\[-0.5ex]
               & d_{t(i), h(j)}, d_{t(i), t(j)} \big)
\end{aligned} & \text{if } i, j \in \mathcal{C}
\end{cases}
\end{equation}
where $d_{ij} = D(i,j)$. Figure \ref{fig: example1} describes how the contraction process works, reducing a 11-node TSP instance into a 5-node sub-TSP instance. The instance has 11 points in the two dimensional Euclidean plane. From the first three solutions in Figure \ref{fig: example1} (a)-(c), by applying a confidence threshold of $\tau = 0.6$, we obtain the set of edges highlighted in blue in Figure \ref{fig: example1} (d). It is important to note that if a higher threshold, say $\tau = 0.9$, were used, the bottom edge would not have been considered. After determining the fixed set of edges, the graph is contracted, as shown in Figure \ref{fig: example1} (e). Note that the two connected nodes in a chain logically point to the head $h(c)$ and tail $t(c)$ endpoints of the path. The chains in the newly constructed distance matrix are logically viewed as a super-node.

Based on \cite{Subqubo}, our proposed method uses a stochastic set of solutions to calculate the edge frequencies instead of using the entire set of available solutions. Rather than calculating $\textit{freq}(e)$ over the entire $\mathcal{P}$, we uniformly sample a smaller subset $\mathcal{S} \subset \mathcal{P}$ of size $N_S < N_I$ at each iteration. The motivation behind this choice, as also explained in \cite{Subqubo}, lies primarily in wanting to avoid staticity and introduce diversification and variety during the main loop iterations, leading to a wider exploration of the solution space.
\subsection{Algorithm Description}
As showed in Algorithm \ref{alg:hybrid_pimc_tsp}, the execution begins with the initialization of the solution pool $\mathcal{P}$ (line 1) through some fast classical heuristic and we retrieve the best solution found. Then the algorithm enters in a ``hybrid'' optimization loop.

The first step of this loop is the refinement of the existing solutions through some classical computation (line 5), as detailed below in Section \ref{subs: 4.2.1}. This step aims to guarantee a progressive improvement in the quality of the pool's solutions and therefore increasing the statistical reliability of the edges selected to be fixed. At line 6, we sample $N_S$ solutions from the pool and analyze their edge frequencies (line 7). If fixed edges are found, the graph is contracted to a subgraph $G_{\textit{sub}}$, and a solution is computed using the PIMC quantum solver (line 9). It is important to note that this particular step of the algorithm can be replaced using a quantum device, for example, a D-Wave quantum annealer, as long as the reduced graph complies with the qubit count and connectivity of the device. Once a new solution is obtained (line 10), we add it to our initial pool and update the global best solution if an improvement is found (line 13). We iterate this process until $K$ consecutive iterations with no improvement of the current solution quality. 
Finally, the algorithm terminates and returns the best tour found.

\begin{algorithm}\label{alg:1}
\caption{Proposed Hybrid Algorithm}
\label{alg:hybrid_pimc_tsp}
\begin{algorithmic}[1] 
\Require $G=(V,E)$: Graph, $D$: distance matrix, $N_I$: initial pool size, $N_S$: sample size, $\tau$: confidence threshold, $K$: max fails allowed, $\gamma$: local search steps 
    \State $\mathcal{P} \gets \text{InitializeClassicalPool}(N_I, \gamma)$ \Comment{SA + 2-opt}
    \State $X_{\textit{best}} \gets \text{GetBestTour}(\mathcal{P})$
    \State $fails \gets 0$
    
    \While{$fails < K$}
        \State $\mathcal{P} \gets \text{RefineSolutions}(\mathcal{P}, \gamma)$ \Comment{Using a classical computer}
        \State $\mathcal{S} \gets \text{SampleSolutions}(\mathcal{P}, N_S)$
        \State $E_{\textit{fixed}} \gets \text{GetConfidentEdges}(\mathcal{S}, \tau)$
        
        
        \State $G_{\textit{sub}}, D_{\textit{sub}} \gets \text{ContractGraph}(G, D, E_{\textit{fixed}})$
        \State $X_{\textit{sub}} \gets \text{PIMC}(G_{\textit{sub}}, D_{\textit{sub}})$ \Comment{Martoňák 2004 scheme}
        \State $X_{\textit{new}} \gets \text{ExpandTour}(X_{\textit{sub}}, E_{\textit{fixed}})$
        
        \State $\mathcal{P} \gets \text{UpdatePool}(\mathcal{P}, X_{\textit{new}})$
        
        \If{$\textit{cost}(X_{\textit{new}}) < \textit{cost}(X_{\textit{best}})$}
            \State $X_{best} \gets X_{\textit{new}}$
            \State $\textit{fails} \gets 0$
        \Else
            \State $\textit{fails} \gets \textit{fails} + 1$
        \EndIf
    \EndWhile
    
    \State \textbf{return} $\textit{cost}(X_{\textit{best}}), X_{\textit{best}}$

\end{algorithmic}
\end{algorithm}
\subsubsection{Classical Initialization and Refinement} \label{subs: 4.2.1}
In the pool initialization step (line 1), a diversified collection of solutions is constructed in a sufficiently short time, producing solutions of good (but possibly sub-optimal) quality. The goal is to obtain a variety of good solutions that allow identifying the edges that most likely belong to the optimal solution. Our pool generation has two main steps:
\begin{enumerate}
    \item The first phase is the generation of random permutations of the cities. This generation utilizes the Fisher-Yates shuffle algorithm \cite{fisher1938statistical, durstenfeld1964algorithm}, which guarantees a uniform distribution over all possible permutations.
    \item The second phase refines each random solution generated in the previous phase via simulated annealing, where each candidate move is a 2-opt transformation. 
\end{enumerate}
An important aspect in this initialization phase is the balance between quality, diversity and efficiency. By starting with random solutions and only making a few small changes we end up with a pool that has many different solutions that preserves good quality, all achieved within a very short computational time. The refinement of the solutions (line 5) is essential to mitigate the structural inefficiencies emerging after the tour expansion and ensuring that the solution achieves global convergence.


\subsubsection{Parameter Dynamics}\label{subsec:parameters}
One drawback of this approach is the number of parameters that need to be tuned in order to control the balance between solution quality and computational cost, the exploration of solutions, and the extent to which the quantum component influences the final solution (the objective being to exploit any quantum capabilities as much as possible). Below, we analyze the main parameters of the algorithm:
\paragraph{Pool dimension $N_I$:} intuitively, generating a large pool of initial solutions increases the probability of accurately capturing the global topological structure of the optimal tour. However a higher $N_I$ can also introduce computational inefficiencies, particularly for large-scale instances. In general, the higher the number of cities, the higher $N_I$ should be.
\paragraph{Sample Size $N_S$:} similarly, with a small $N_S$ value we have a higher variance in the estimated frequencies, hence a more variable behavior of the algorithm. With a high value moving towards $N_I$, the frequencies stabilize and the behavior is more deterministic.
\paragraph{Threshold $\tau$:} selecting an appropriate threshold value depends on several factors, the most significant one is the instance dimension. As shown in Table \ref{tab:impact_threshold}, high threshold values can result in a conservative contraction, where only nearly universally present edges are fixed. Conversely, when the threshold values are lower, the original instance is contracted aggressively, which carries a high risk of fixing sub-optimal edges that do not belong to the global optimum.
\paragraph{Local search steps $\gamma$:} this parameter controls the weight of the classical computation in both the initialization and the refinement of the solution. Low values of this parameter reduce the impact of the classical solver, leading to a greater involvement of the quantum component. High values introduce a higher computational cost, but yield more stable and significant edge frequencies.


%% file: Sections/ExperimentalSetup.tex
\section{Experimental Setup}\label{sec:expsetup}
To rigorously evaluate the efficacy of our proposed hybrid quantum-classical framework, we performed an experimental evaluation focused on parameter sensitivity and comparative benchmarking.
Our evaluation addresses three primary objectives:
\begin{itemize}
    \item to quantify the impact of the three principal parameters on the final solution quality: confidence threshold $\tau$, pool dimension $N_I$ and sample size $N_S$; 
    \item to benchmark the overall performance of the approach across multiple selected TSP instances against a classical method (Google's OR-Tools), using the optimal parameter configuration identified in the first stage of our experiment;
    \item to evaluate our Hybrid Algorithm, where PIMC is replaced by the D-Wave QPU, against the D-Wave hybrid solver.
\end{itemize} 
We use instances from the dataset provided in the TSPLIB library \cite{tsplib}, the standard library of TSP benchmark instances.
The performance of the algorithm is quantified using several metrics:
\begin{itemize}
    \item the \textit{Optimality Gap} quantifies the relative percentage deviation of the obtained results from the known optimum; 
    \item execution times required for the algorithm to reach convergence. Specifically, we measure the overall execution time and the total time spent on the PIMC simulation. For the experiments performed with D-Wave, we took into consideration the values of \textit{annealing time};
    \item the \textit{Compression Rate}, which indicates the degree of dimensionality reduction achieved during the extraction phases across different parameter configuration. 
\end{itemize} 
Our Algorithm \ref{alg:hybrid_pimc_tsp} was implemented in C++, and all the experiments were executed on a AMD Ryzen 7 laptop, running Ubuntu 25.04 with 16 GB of RAM; the source code is publicly available \cite{source}.

%% file: Sections/Results.tex
\section{Results and Discussion} \label{sec:results}
\begin{table*}[t]
\centering
\caption{Impact of the confidence threshold ($\tau$) on the compression rate, solution quality, and computational time across three TSPLIB instances. Results are averaged over 100 independent runs, with $N_I = 100$, $N_S = 50$, $\gamma = 1$ and $K = 3$ held constant.}
\label{tab:impact_threshold}
\begin{tabular}{@{}llccccc@{}}
\toprule
\textbf{Instance ($N$)} & \textbf{Threshold ($\tau$)} & \textbf{$N_{sub}$ (Nodes)} & \textbf{Compression Rate (\%)} & \textbf{Optimality Gap (\%)} & \textbf{Total Time (s)} & \textbf{PIMC Time (s)}\\ \midrule

\textbf{\texttt{berlin52}}  & 1.00 & 49.17 $\pm$ 0.65 & 5.00 $\pm$ 0.57  &  0.08 $\pm$ 0.26 & 1.68 $\pm$ 0.35 & 1.65 $\pm$ 0.35 \\
                            & 0.90 & 42.24 $\pm$ 1.17 & 17.7 $\pm$ 2.00  &  0.24 $\pm$ 0.47 & 1.35 $\pm$ 0.32 & 1.32 $\pm$ 0.32 \\
                            & 0.75 & 31.28 $\pm$ 1.05 & 37.91 $\pm$ 0.93 &  0.92 $\pm$ 0.84 & 0.88 $\pm$ 0.20 & 0.85 $\pm$ 0.19 \\
                            & 0.50 & 8.32  $\pm$ 0.93 & 83.83 $\pm$ 0.90 &  1.37 $\pm$ 0.99 & 0.09 $\pm$ 0.02 & 0.06 $\pm$ 0.02 \\ \midrule

\textbf{\texttt{pr439}}     & 1.00 & 417.81 $\pm$ 3.11 & 4.83  $\pm$ 0.71 & 0.69 $\pm$ 0.15  & 52.75 $\pm$ 15.31 &  52.35 $\pm$ 15.23 \\
                            & 0.90 & 335.83 $\pm$ 7.01 & 23.50 $\pm$ 1.60 & 0.88 $\pm$ 0.17  & 31.69 $\pm$ 6.44  &  31.31 $\pm$ 6.39  \\
                            & 0.75 & 201.88 $\pm$ 6.55 & 54.01 $\pm$ 1.49 & 1.29 $\pm$ 0.27  & 10.85 $\pm$ 1.38  &  10.49 $\pm$ 1.35  \\
                            & 0.50 & 51.12  $\pm$ 2.93 & 88.36 $\pm$ 0.67 & 2.60 $\pm$ 0.64  & 1.90  $\pm$ 0.41  &  1.46  $\pm$ 0.33  \\ \midrule

\textbf{\texttt{pr1002}}    & 1.00 & 976.63 $\pm$ 2.58 & 2.23 $\pm$ 0.26 & 1.64 $\pm$ 0.21 & 258.38 $\pm$ 78.02 & 256.89 $\pm$ 77.81\\
                            & 0.90 & 860.91 $\pm$ 9.69 & 14.08 $\pm$ 0.97 & 1.88 $\pm$ 0.19 & 173.25 $\pm$ 40.53 & 171.86 $\pm$ 40.39\\
                            & 0.75 & 621.23 $\pm$ 9.77 & 38.00 $\pm$ 0.98 & 2.55 $\pm$ 0.31 & 77.37 $\pm$ 9.91 & 76.03 $\pm$ 9.82 \\
                            & 0.50 & 184.90 $\pm$ 5.86 & 81.55 $\pm$ 0.59 & 5.69 $\pm$ 0.52 &9.53 $\pm$ 1.25 & 8.07 $\pm$ 1.13\\ \bottomrule
\end{tabular}
\end{table*}


\subsection{Threshold Impact}
Table \ref{tab:impact_threshold} illustrates the influence of the confidence threshold $\tau$ on the compression rate, solution quality and computational performance. A primary observation is the significant reduction in problem dimensionality ($N_{sub}$) and execution time as $\tau$ is lowered. While an aggressive contraction policy effectively reduces the graph, it introduces a measurable increase in the optimality gap, highlighting a trade-off between computational performance and solution quality.  

In the \texttt{berlin52} instance, at a threshold of $\tau = 1.00$, the graph is compressed by a modest 5\%, resulting in a sub-problem of approximately 49 nodes. Given the constraints of current quantum hardware and simulation environments, a problem of such size remains exceedingly difficult to map and solve efficiently. Conversely, setting $\tau = 0.50$ achieves an 83\% average reduction, reducing the instance to roughly 8 nodes, which falls within the scale supported by current quantum devices. However, this aggressive pruning of the search space increases the probability of fixing sub-optimal edges, thereby degrading the final solution quality, as reflected by the increase in the optimality gap. The threshold $\tau$ must therefore be chosen as a compromise between compression and accuracy. 

Aggressive graph reduction also yields substantial benefits in terms of computational efficiency, particularly regarding the quantum simulation phase. In the \texttt{pr1002} instance, the total execution time for $\tau = 0.50$ is dramatically lower than that required for $\tau = 1.00$. Furthermore, the data reveal a near-identical relationship between Total Time and PIMC Time. This indicates that the ``classical'' overhead, which includes initial pool generation, solution refinement and graph contraction, contributes only marginally to the overall runtime compared to the computational intensity of the PIMC simulation. Consequently, the PIMC phase remains the primary bottleneck, and its duration is directly governed by the effectiveness of the threshold-driven dimensionality reduction.

\subsection{Pool Size and Sample Size Impact}

Table \ref{tab:impact_pool_sample} illustrates the influence of initial solution pool size $N_I$ and the sample size $N_S$ on the optimality gap and computational overhead. These experiments were conducted using a fixed confidence threshold $\tau = 0.80$, with $K = 3$ and $\gamma = 1$. The tested values for $N_S$ were set at 20\%, 50\% and 100\% of the corresponding pool size $N_I$.

The main outcome is that the best solution quality across all instances was consistently achieved with the configuration $N_I = 500$ and $N_S = 250$. This indicates that a sufficiently large pool of candidate solutions is beneficial, since it provides a more reliable basis for identifying stable edges during graph contraction phase. Interestingly, the data suggests that setting $N_S = N_I$ for large pools does not always yield better results. When the pool is sufficiently large ($N_I = 500$), evaluating every solution makes the edge selection process overly deterministic. In contrast, sampling a subset appears to facilitate a more exhaustive exploration of the solution space. This is precisely what is observed for the configuration $N_I = 500$, where $N_S = 250$ consistently yields better results than $N_S = 500$.

However, a reverse trend is observed for the larger instances when the pool size is small. In the \texttt{pr439} and \texttt{pr1002} instances, the choice $N_I = 50$ produces better results when the full pool is considered, that is, when $N_S = 50$. In this scenario, the pool is too small to provide strong statistical significance for instances involving several hundred or more than one thousand nodes. Extracting a tiny sample set, such as $N_S = 10$ or $N_S = 25$, introduces substantial ``statistical noise'', leading to less reliable frequency estimates for edge selection. Consequently, using the entire pool is preferable. This is also evidenced by the \texttt{berlin52} instance: since the problem is much smaller, a pool of 50 solutions is already reasonably representative, and the advantage of taking $N_S < N_I$ remains visible.

A final observation concerns the contribution to the execution time. The data reveal that the classical computational overhead is remarkably low. For example, in the \texttt{pr1002} instance, increasing the pool size from 50 to 500 adds only a few seconds to the total runtime. As noted in the previous section, the dominant computational cost remains the PIMC, which could be entirely replaced by execution on quantum hardware, offering the potential for an exponential decrease in computational complexity.

\begin{table*}[t]
\centering
\caption{Impact of the initial pool size ($N_I$) and sample size ($N_S$) on solution quality and computational time across three TSPLIB instances. The values of $N_S$ correspond to 20\%, 50\% and 100\% of $N_I$. The confidence threshold, classical refinement steps and other parameters are held constant (i.e., $\tau = 0.8$, $K = 3$, $\gamma = 1$).}
\label{tab:impact_pool_sample}
\begin{tabular}{@{}llcccc@{}}
\toprule
\textbf{Instance ($N$)} & \textbf{Pool Size ($N_I$)} & \textbf{Sample Size ($N_S$)} & \textbf{Optimality Gap (\%)} & \textbf{Total Time (s)} & \textbf{PIMC Time (s)} \\ \midrule

\textbf{\texttt{berlin52}}  & 50 & 10  &0.79 $\pm$ 1.00&1.06 $\pm$ 0.25&1.05 $\pm$ 0.25 \\
                            & 50 & 25  &0.85 $\pm$ 0.77&1.31 $\pm$ 0.29&1.29 $\pm$ 0.29\\                    
                            & 50 & 50 &0.87 $\pm$ 0.90&1.39 $\pm$ 0.33&1.38 $\pm$ 0.33\\ \cmidrule(l){2-6}
                            & 500 & 100  &0.33 $\pm$ 0.48&1.51 $\pm$ 0.29&1.40 $\pm$ 0.27\\
                            & 500 & 250 &\underline{0.15 $\pm$ 0.37}&1.58  $\pm$ 0.32&1.46 $\pm$ 0.30 \\
                            & 500 & 500 &0.24 $\pm$ 0.43&1.54 $\pm$  0.30&1.42 $\pm$ 0.28 \\ \midrule

\textbf{\texttt{pr439}}     & 50 & 10  &1.41 $\pm$ 0.31&9.75 $\pm$ 1.48&9.59 $\pm$ 1.46\\
                            & 50 & 25  &1.21 $\pm$ 0.25&11.37 $\pm$ 1.72&11.21 $\pm$ 1.70\\
                            & 50 & 50 &1.16 $\pm$ 0.24&11.96 $\pm$ 1.69&11.80 $\pm$ 1.67\\ \cmidrule(l){2-6}
                            & 500 & 100  &1.16 $\pm$ 0.22&15.03 $\pm$ 2.31&13.46 $\pm$ 2.11\\
                            & 500 & 250 &\underline{1.13 $\pm$ 0.21}&15.40 $\pm$ 2.46&13.84 $\pm$ 2.25\\
                            & 500 & 500 &1.14 $\pm$ 0.24&14.96 $\pm$ 2.07&13.45 $\pm$ 1.90\\ \midrule

\textbf{\texttt{pr1002}}    & 50 & 10  &2.90 $\pm$ 0.33&52.30 $\pm$ 8.22&51.74 $\pm$ 8.16\\
                            & 50 & 25  &2.40 $\pm$ 0.30&62.11 $\pm$ 7.55&61.56 $\pm$ 7.51 \\
                            & 50 & 50 &2.26 $\pm$ 0.31&65.73 $\pm$ 9.42&65.20 $\pm$ 9.38\\ \cmidrule(l){2-6}
                            & 500 & 100  &2.17 $\pm$ 0.27&73.62 $\pm$ 9.55 &68.69 $\pm$ 9.18 \\
                            & 500 & 250 &\underline{2.17 $\pm$ 0.28}&76.44 $\pm$ 9.91&71.48 $\pm$ 9.54\\
                            & 500 & 500 &2.20 $\pm$ 0.24&79.40 $\pm$ 11.23&74.35 $\pm$ 10.82\\ \bottomrule
\end{tabular}
\end{table*}
\subsection{Overall Performance}
Table \ref{tab:overall_performance} summarizes the performance of our approach across a diverse set of TSPLIB instances (digits at the end of an instance's name identify the number of cities, {\em e.g.}, \texttt{pr1002} has 1,002 cities). We used the parameter configuration selected from the previous sensitivity analysis as the most effective compromise between solution quality and computational cost: $\tau = 0.75$, $N_I=500$ and $N_S = 250$. Also, we benchmarked our results against Google's OR-Tools (Vehicle Routing module) by setting the same time limit obtained by our approach. We configured the \textit{Path Cheapest Arc} as the first solution strategy and \textit{Guided Local Search} as the metaheuristic.

For the smaller instances \texttt{burma14} and \texttt{ulysses22}, our algorithm  identified the global optimum in every trial. Notably, the resulting sub-problem sizes demonstrate that our hybrid pre-processing effectively reduces low-dimensional problems to a scale that is directly mappable onto current QPUs.

\begin{table*}[t]
\centering
\caption{Overall performance of the proposed Hybrid Algorithm across selected TSPLIB instances. The evaluation is performed using the suitable parameter configuration derived from the sensitivity analysis ($\tau = \text{0.75}$, $N_I = \text{500}$, $N_S = \text{250}$, $\gamma = 1$ and $K = 3$). Results are averaged over 100 independent runs. For the specific large instance \texttt{rl11849} we employed $\tau = 0.3$ and averaged the results over 10 runs.}
\label{tab:overall_performance}
\begin{tabular}{@{}l|ccccc|cc@{}}
\toprule
 & \multicolumn{5}{c|}{\textbf{Proposed Hybrid Algorithm}} & \multicolumn{2}{c}{\textbf{OR-Tools}} \\ \cmidrule(lr){2-6} \cmidrule(l){7-8}
\textbf{Instance}  & \textbf{Avg. $N_{sub}$} & \textbf{Avg. Opt. Gap (\%)} & \textbf{Best Opt. Gap (\%)} & \textbf{Total Time (s)} & \textbf{PIMC (s)} & \textbf{Opt. Gap (\%)} & \textbf{Time (s)} \\ \midrule
\texttt{burma14}                           & 4.93  $\pm$ 0.35  & 0.00 & 0.00 & 0.02 $\pm$ 0.0 & 0.01 $\pm$ 0.0 & 0.00 & 0.01 \\
\texttt{ulysses22}                         & 10.56 $\pm$ 0.26  & 0.00 & 0.00 & 0.08 $\pm$ 0.0 & 0.05 $\pm$ 0.0 & 0.96 & 1 \\
\texttt{berlin52}                          & 34.02 $\pm$ 0.71  & 0.34 $\pm$ 0.56 & 0.00   & 0.97 $\pm$ 0.17 & 0.87 $\pm$ 0.15 & 6.05 & 1 \\
\texttt{pr264}                             & 78.76 $\pm$ 8.05  & 0.75 $\pm$ 0.67 & 0.14   & 4.36 $\pm$ 0.54 & 3.42 $\pm$ 0.38 & 4.80 & 5 \\
\texttt{pr439}                             & 204.33 $\pm$ 5.44 & 1.30 $\pm$ 0.26 & 0.80   & 12.24 $\pm$ 1.38 & 10.49 $\pm$ 1.22 & 6.44 & 12 \\
\texttt{pr1002}                            & 634.26 $\pm$ 7.39 & 2.50 $\pm$ 0.32 & 1.79   & 84.55 $\pm$ 7.64 & 78.04 $\pm$ 7.36 & 4.16 & 85 \\
\texttt{pr2392}                            & 1448.22 $\pm$ 32.07 & 3.95 $\pm$ 1.34 & 3.01   & 694.66 $\pm$ 85.97 & 652.74 $\pm$ 83.64 & 4.54 & 650 \\ \cmidrule(l){2-8}
\texttt{rl11849}                          & 1146.36  $\pm$ 11.42    & 7.36 $\pm$ 0.24   & 7.11   &   2186.94 $\pm$ 251.38 & 650.38 $\pm$ 142.60 & 14.58 & 2263\\ \bottomrule
\end{tabular}
\end{table*}

\begin{table*}[t]

\centering
\caption{Performance comparison between the D-Wave BQM Hybrid Solver and the proposed Hybrid Algorithm using the D-Wave Advantage\_system 4.1 QPU. Results are averaged over 10 independent runs.}
\label{tab:dwave_hybrid_comparison}
\begin{tabular}{@{}l|cc|ccccc@{}}
\toprule
 & \multicolumn{2}{c|}{\textbf{D-Wave BQM Hybrid Solver}} & \multicolumn{5}{c}{\textbf{Proposed Hybrid Algorithm with D-Wave QPU}} \\ \cmidrule(lr){2-3} \cmidrule(l){4-8}
\textbf{Instance} & \textbf{Opt. Gap (\%)} & \textbf{QPU Time (ms)} & \textbf{Avg. $N_{sub}$} & \textbf{Opt. Gap (\%)} & \textbf{Tot. Time (ms)} & \textbf{Clas. Time (ms)} & \textbf{QPU Time (ms)} \\ \midrule
\texttt{burma14}  & 2.26 $\pm$ 0.89 & 103.73 $\pm$ 15.91 & 4.02 $\pm$ 0.23 & 1.75 $\pm$ 1.88 & 11549.90 $\pm$ 2867.69 & 55.80 $\pm$ 12.10 & 97.62 $\pm$ 23.42 \\
\texttt{ulysses16}  & 5.02 $\pm$ 1.58 & 99.45 $\pm$ 0.36  & 3.94 $\pm$ 0.66 & 1.20 $\pm$ 0.73 & 12924.75 $\pm$ 5278.47 & 59.36 $\pm$ 48.52 & 92.75 $\pm$ 25.41 \\
\texttt{ulysses22}  & 16.88 $\pm$ 3.36 & 101.40 $\pm$ 0.84 & 5.05 $\pm$ 0.69 & 2.57 $\pm$ 1.63 & 15067.80 $\pm$ 4872.35 & 137.80 $\pm$ 54.10 & 130.59 $\pm$ 40.79 \\ \bottomrule
\end{tabular}
\end{table*}

As the problem size increases, the algorithm maintains high solution quality, with the mean optimality gap remaining below 4\% even for the a large instance like \texttt{pr2394}. This indicates that the proposed contraction strategy is able to reduce the effective problem size substantially without causing a severe degradation in solution quality. The proximity between the mean and the best optimality gap across all instances suggests our technique is highly robust and not overly sensitive to the stochastic nature of the initial pool generation. For instance, in \texttt{pr264}, the best observed gap was as low as 0.14\%, indicating that our technique is capable of reaching near-optimal configuration with high reliability.

To investigate the limits of our technique, we tested the large-scale instance \texttt{rl11849}. Given the size of the problem (11,849 cities), we intentionally reduced the compression threshold to $\tau = 0.30$ to obtain a smaller sub-problem. This led to an average graph reduction of about 90\%. At this scale, however, the main computation bottleneck is no longer the PIMC solver. A lower threshold $\tau$ allows the contracted subproblem to be substantially smaller, which reduces the time spent in the PIMC optimization. On the other hand, the classical stages of the algorithm introduce a significant overhead (almost 70\% of the total runtime was spent in the classical stages: 33\% in pool initialization and 36\% in the refinement step). The method still delivers a good solution quality, with an optimality gap slightly above 7\% (OR-Tools returns a 14\% gap). As result, an appropriate choice of $\tau$ can effectively reduce the cost of PIMC, but it does not alleviate the classical overhead, which eventually becomes the dominant bottleneck. 

Overall, our obtained results support the effectiveness of our method: it achieves exact solutions on small instances, maintains low optimality gaps on medium and large-scale problems, and benefits from substantial instance compression, thereby moving the reduced subproblems closer to the size range that can be handled by current quantum annealers. Nevertheless, the results on the largest benchmark also reveal a limitation of the approach, where the computation overhead of the classical stages can eventually prevail over the benefit of solving a smaller contracted sub-problem. However, this is reasonable as we are essentially requesting to reduce the size of the problem that is passed to the quantum device, thereby waving the potential gains of a quantum solution.

\subsection{D-Wave Hardware Results}
Table \ref{tab:dwave_hybrid_comparison} shows how our algorithm performs on the D-Wave Advantage 4.1. We intentionally omitted the pool refinement phase to prevent the trivial identification of the optimal tour during the initial steps. The results are benchmarked against those obtained by D-Wave's BQM hybrid solver. In both cases, we employed the same QUBO defined in Eq.~\eqref{qubo}. The Lagrange multipliers $\lambda_1$ and $\lambda_2$ were set to $0.8 \times \textit{max}\_\textit{dist}$, where \textit{max}$\_$\textit{dist} is the maximum distance between any two nodes in the instance. For the QPU, we set the annealing time to $20 \mu s$ with 25 number of reads, while the time limit imposed on the D-Wave BQM hybrid solver is 3 seconds. The \texttt{DWaveCliqueSampler} was employed to minor-embed the QUBO model into the Pegasus topology. Regarding the algorithm parameters, we used a pool size $N_I = 100$, a sample size $N_S = 50$ and local search steps $\gamma = 1$. For \texttt{burma14} and \texttt{ulysses16} we used a threshold $\tau = 0.4$, whereas for \texttt{ulysses22} we had to use $\tau = 0.3$ in order to achieve a significant reduction for direct execution on the QPU. For these experiments, we aimed to evaluate the impact of quantum processing in solving the instances. It is worth noting that the total execution time (which includes minor-embedding, network latency, and QPU queue wait times) is significantly higher than the classical processing time of our algorithm (which excludes the overhead required by the quantum hardware).
The results indicate that by significantly reducing the initial instance, high-quality results can be achieved within competitive QPU times, even outperforming the average optimality gap of the D-Wave BQM Hybrid solver. Taking \texttt{ulysses22} as an example, we can observe that our approach is able to achieve an average optimality gap of 2.57$\%$, compared to D-Wave's 16.88$\%$, by solving sub-instances of 5 nodes on average. 

%% file: Sections/Conclusions.tex
\section{Conclusion}

We introduced a hybrid solution framework for the TSP that combines graph contraction based on edge frequency with a PIMC optimization stage. The proposed method is motivated by the observation that direct quantum formulations of the TSP remain severely constrained by current hardware limitations, whereas an adaptive contraction strategy can reduce the effective problem size while preserving the most stable structural information contained in a pool of candidate tours.
The experimental analysis shows that the method achieves a favorable balance between solution quality and computational efficiency. For small benchmark instances, the algorithm recovers the optimal solution, while for medium-large scale instances it maintains low optimality gaps. At the same time, the contraction mechanism substantially decreases the problem size, bringing them closer to the scale supported by current quantum annealers.
Our experiments make clear that the primary bottleneck is the PIMC quantum annealing step. While currently implemented by classical simulation, the direct execution of this step on a quantum annealer could drastically reduce the computational cost.